\documentclass[twocolumn]{revtex4}
\usepackage[utf8]{inputenc}
\usepackage{graphicx}

\begin{document}

\title{Regimes of optical transparency and instabilities of collinear dielectric ferromagnetic materials in the presence of the dynamic magnetoelectric effect}

\author{Pavel A. Andreev}
\email{andreevpa@physics.msu.ru}
\affiliation{Department of General Physics, Faculty of physics, Lomonosov Moscow State University, Moscow, Russian Federation, 119991.}

\date{\today}

\begin{abstract}
The contribution of the polarization associated with the noncollinear parts of spins in the dielectric permeability tensor of multiferroic materials is considered.
As the equilibrium state, we consider the systems of parallel spins,
so we have zero equilibrium polarization.
Dynamical polarization appears due to the spin evolution via the magnetoelectric coupling.
The regime of frequencies,
where the refractive index goes to 1 is found for the high
(in comparison with the characteristic frequency of the anisotropic energy or the cyclotron frequency)
left/right circularly polarized electromagnetic wave propagating parallel to the equilibrium spin direction.
Moreover, the signs of the imaginary part of the dielectric permeability, the refractive index, and the frequency
show the instability of the parallel spin configuration
at the propagation of the linearly polarized electromagnetic wave due to the effective magnetoelectric interaction.
\end{abstract}


\maketitle


\section{Introduction}

The collinear order of spins in the ferromagnetic and antiferromagnetic materials is caused by the Coulomb exchange interaction,
which can be described by the Heisenberg Hamiltonian.
The noncollinear order of spins in other magnetically ordered materials corresponds to the instability of the collinear state
due to some interaction between ions.
The potential energy of the
Dzyloshinskii--Moriya interaction
has its minimum at the perpendicular orientation of neigbouring spins.
However,
the Dzyloshinskii--Moriya interaction leads to the stable dispersion dependence of the spin waves $\omega(\textbf{k})$
\cite{Camley SSR 23}.
It is well-known that
the Dzyloshinskii--Moriya interaction gives the dependence of the frequency on the projection of the wave vector,
in contrast with the Heisenberg exchange interaction giving dependence on the square of the wave vector.
Therefore, it is necessary to identify the interaction leading to the instability of the collinear spin order,
which manifests itself in the signature of the imaginary part of frequency.

In this paper we consider
the magnetoelectric effect in ferromagnetic multiferroic materials with the ferroelectricity of spin origin.
While the considered equilibrium condition is constructed of parallel spins,
we focus our attention on the polarization of the medium associated with noncollinear spins
\cite{Pyatakov UFN 12}, \cite{Khomskii JETP 21}.
In the considered regime,
the noncollinear part of spins appears due to the spin perturbations.
Nature of the static and dynamic of the magnetoelectric effect associated with the spin-orbit interaction.
The existence of polarization of spin origin leads to interaction of the external electric field with the polarization and
interaction of perturbations of the inner electric field with the polarization.
We include this interaction in the Landau--Lifshitz--Gilbert equation
(like it is done in Ref. \cite{Risinggard SR 16} for the static electric field).

The problem of instability of the collinear order of spins
in the antiferromagnetic easy-plane multiferroics
due to the magnetoelectric interaction is addressed in
Ref. \cite{ZvezdinMukhin JETP L 09}.
Authors find the possibility of the zero frequency solution for the small amplitude perturbations
(at the nonzero wave vector).
It can be interpreted as a motionless spiral structure of spins,
where the wave vector gives the spatial period of this structure.
This structure looks like the usual small-amplitude spin wave at the fixed moment in time,
where we see the small deviation of spins from the equilibrium direction.
So, no large rotation of the spin direction over the distance can be found here.
No explicit argument about the imaginary part of this solution
(related to the Gilbert damping) is given.
So, this zero-frequency spiral can relax towards the collinear state.
Perturbations of the electric and magnetic fields are also considered in
Ref. \cite{ZvezdinMukhin JETP L 09},
but no assumptions are described.
However, we can conclude that some assumptions are made
since authors obtain two spin-wave solutions
(it is related to the consideration of the antiferromagnetic with two ferromagnetic sublattices),
while no electromagnetic wave branch of the spectrum is explicitly presented.
Our goal is to identify the instability of the collinear order of spins
via the derivation of the imaginary part of frequency, showing the
growth
of the amplitude of perturbations in spite of the presence of the Gilbert damping.
This instability also reveals itself in the sign of the imaginary part of the elements of
the dielectric permeability tensor and the refractive index.

It is demonstrated in
Ref. \cite{ZvezdinMukhin JETP L 09},
that the magnetoelectric interaction gives the effect similar to
the Dzyloshinskii-Moriya interaction,
where the spectrum differs for the waves propagating along and against the chosen direction.
In other words, there is dependence of the frequency on the projection of the wave vector.
This is one of the distinctive features of the Dzyloshinskii-Moriya interaction and
the effective magnetoelectric interaction.
In this paper, we are focused on the systematic analysis of the most distinctive features
of the dielectric permeability, the refractive index, and the dispersion dependence
for the frequency
of the spin and electromagnetic waves propagating in the dielectric mediums
in the presence of the effective magnetoelectric interaction.
Particularly, we are interested in the instability of the equilibrium collinear state towards the possibility of noncollinear structure formation.

This paper is organized as follows.
In Sec. II we discuss the form of the electric polarization-spin relation
and the contribution of the polarization in the Landau--Lifshitz--Gilbert equation.
In Sec. III we present the background model for the derivation of the dielectric permeability.
In Sec. IV we analyze the waves propagating perpendicular to the anisotropy axis.
In Sec. V we analyze the waves propagating parallel to the anisotropy axis.
In Sec. VI a brief summary of obtained results is presented.

\section{Macroscopic spin polarization of the spin origin}

Phenomenological macroscopic analysis based on the symmetries is the highly useful tool in condensed matter physics.
Particularly, the magneto-electric effect in the multiferroic materials is related to breaking both the spatial
inversion and the time reversal symmetries.
It happens due to the simultaneous presence of the ferroelectricity and the ferromagnetism in the multiferroic materials
\cite{Mostovoy npj 24}.
However, it is necessary to understand the microscopic mechanisms leading to the appearance of the magnetoelectric coupling
like it is described in Ref. \cite{AndreevTrukh PS 24}.

Macroscopic polarization can be defined as the quantum average of the sum of the single particle operators:
\begin{equation}\label{MFMemf P def} \textbf{P}(\textbf{r},t)=
\int \Psi_{S}^{\dagger}(R,t)\sum_{i}\delta(\textbf{r}-\textbf{r}_{i})
(\hat{\textbf{d}}_{i}\Psi(R,t))_{S}dR. \end{equation}
Here the
electric dipole moment
(edm)
operator $\hat{\textbf{d}}_{i}$ is the relative shift of the magnetic ion from the nonmagnetic ion $\Delta\textbf{r}_{i}$,
where we see the relative shift of ions from positions with coincidence of the center of "mass/charge" of positive and negative charges $\hat{\textbf{d}}_{i}=q_{i}\Delta\textbf{r}_{i}$.
In general case, it can be written as $\hat{\textbf{d}}_{i}=q_{i}\textbf{r}_{i}$,
but such a form requires the explicit account of the positions and dipole moments of the nonmagnetic ions.

The Dzyloshinskii--Moriya interaction can
lead to the noncollinear spin structures.
If the noncollinear spin structure is formed in the magnetic samples,
it can lead to the formation of the electric dipole moment due to the shifts of the nonmagnetic ions relatively to the magnetic ions.
The spin current model is derived from the analysis  of the momentum balance equation of the set of quantum hydrodynamic equations
\cite{AndreevTrukh PS 24}.
This derivation shows
that the
Heisenberg exchange interaction
forming the magnon spin current leads to the polarization,
which corresponds to the following microscopic operator of edm
\cite{Tokura RPP 14}
\begin{equation}\label{MFMemf edm def simm}
\textbf{d}_{ij}= \alpha_{ij}[\textbf{r}_{ij}\times[\textbf{s}_{i}\times\textbf{s}_{j}]]. \end{equation}
Presented form of
edm of three ions
(two magnetic ions and one nonmagnetic ligand ion)
is presented in the  form symmetric relatively to two magnetic ions.

Each magnetic ion gives the partial contribution to the formation of
the edm with each magnetic neighbor.
We can shift our focus from the pair of ions forming
edm, to each magnetic ion instead:
\begin{equation}\label{MFMemf edm operator Mod}
\hat{\textbf{d}}_{i}=\sum_{j\neq i}
\alpha_{ij}(r_{ij})[\textbf{r}_{ij}\times[\hat{\textbf{s}}_{i}\times\hat{\textbf{s}}_{j}]], \end{equation}
where we introduce function
$\alpha_{ij}(r_{ij})=\alpha_{ij}$
if $r< a_{eff}$
$\alpha_{ij}(r_{ij})=0$ for $r> a_{eff}$,
which allows to ensure that magnetic ions located close to the chosen magnetic ion give the contribution in
the edm formation.

The quantum hydrodynamic method
\cite{AndreevTrukh PS 24}, \cite{AndreevTrukh EPJ B 24}, \cite{AndreevTrukh JETP 24}, \cite{Andreev Ph B 24}, \cite{Andreev 23 11}
allows to find the macroscopic polarization of the medium from the operator
(\ref{MFMemf edm operator Mod})
\begin{equation}\label{MFMemf P def expanded} \textbf{P}(\textbf{r},t)=
\frac{1}{3}g_{(\alpha)}[(\textbf{S}\cdot\nabla)\textbf{S}-\textbf{S}(\nabla\cdot \textbf{S})], \end{equation}
where
$g_{(\alpha)}=\int \xi^{2}\alpha(\xi) d\mbox{\boldmath $\xi$}$.
This result correspond to \cite{Sparavigna PRB 94},
\cite{Mostovoy PRL 06}
and \cite{Dong AinP 15} (see p. 533).

We find the following contribution in the
Landau--Lifshitz--Gilbert equation
$$\partial_{t}S^{\alpha}\mid_{\textbf{E}}=
-\sigma
\varepsilon^{\alpha\beta\gamma}S^{\beta}
\biggl[2E^{\mu}\partial^{\gamma}S^{\mu}$$
\begin{equation}\label{MFMemf evol S under infl of E fin 00}
-2E^{\gamma}(\nabla\cdot\textbf{S})
+S^{\mu}\partial^{\gamma}E^{\mu} -(\textbf{S}\cdot\nabla)E^{\gamma}\biggr].\end{equation}
Derivation of the contribution of the magnetoelectric coupling in the Landau--Lifshitz--Gilbert equation
(\ref{MFMemf evol S under infl of E fin 00}) is made within the quantum hydrodynamic method
\cite{AndreevTrukh PS 24}, \cite{AndreevTrukh EPJ B 24}, \cite{AndreevTrukh JETP 24}, \cite{Andreev Ph B 24}, \cite{Andreev 23 11}
as the derivation of polarization (\ref{MFMemf P def expanded}).
We apply notation $\sigma\equiv -\frac{1}{3}g_{(\alpha)}=\gamma_{0}\gamma^{2}>0$
in comparison with parameter $\gamma_{0}$ used in Ref. \cite{Risinggard SR 16}.

Equation (\ref{MFMemf evol S under infl of E fin 00}) is one of the forms for the torque acting on the spin density due to the magnetoelectric coupling.
It can also be presented in the following form
\begin{equation}\label{MFMemf evol S under infl of E fin 01}
\partial_{t}S^{\alpha}\mid_{\textbf{E}}=-\sigma
\varepsilon^{\beta\gamma\delta}\varepsilon^{\delta\mu\nu}\varepsilon^{\nu\alpha\sigma}
[ S^{\mu}S^{\sigma}\partial^{\gamma}E^{\beta} +2E^{\beta}S^{\sigma}\partial^{\gamma}S^{\mu}].
\end{equation}
Convolution on index $\delta$ of the first two Levi-Civita symbols leads to equation
(\ref{MFMemf evol S under infl of E fin 00}).
Convolution on index $\nu$ of the last two Levi-Civita symbols leads to the expression
$$\partial_{t}S^{\alpha}\mid_{\textbf{E}}=-\sigma
\biggl[-S^{2}[\nabla\times\textbf{E}]^{\alpha}
+\varepsilon^{\alpha\beta\gamma}E^{\beta}\partial^{\gamma}S^{2}$$
\begin{equation}\label{MFMemf evol S under infl of E fin 02}
+S^{\alpha}(\textbf{S}\cdot [\nabla\times\textbf{E}])
-2\varepsilon^{\beta\gamma\delta}S^{\beta}E^{\gamma}\partial^{\delta} S^{\alpha}\biggr].
\end{equation}
We finally find three equivalent results
(\ref{MFMemf evol S under infl of E fin 00}), (\ref{MFMemf evol S under infl of E fin 02}).
Equations (\ref{MFMemf evol S under infl of E fin 00}), (\ref{MFMemf evol S under infl of E fin 01}),
(\ref{MFMemf evol S under infl of E fin 02})
are found in the mean-field approximation.

\subsection{On the estimation of constant $\sigma$}

There are several estimations for the constant of the magneto-electric coupling $\sigma$
(see \cite{Risinggard SR 16}
(one value in Table 1 and another is given by the second equation on p.7),
and Refs. \cite{Sparavigna PRB 94}, \cite{AndreevTrukh PS 24},
\cite{Logginov JETP L 07}).

A simple estimation of the polarization of systems of ions can be considered as the shift of
the singly ionized atom on 1 angstrem.
In the SI units,
it gives us
$P=0.2$
C/m$^2$,
while we work in CGS units and find
$P=5\times10^{4}$
CGS
(at the concentration $n_{0}=10^{22}$cm$^{-3}$).

The generalized spin-current model developed in Refs.
\cite{AndreevTrukh PS 24}, \cite{AndreevTrukh JETP 24}
gives the representation of the constant $\sigma$ in terms of the constant of the exchange interaction $A$ described by the Heisenberg Hamiltonian.
This generalization of the spin-current model presents the systematic relation between the microscopic scale and the macroscopic scale
and describes the microscopic nature of the polarization formation in two regimes
of the polarization formation (for the multiferroics of spin origin).

Ref. \cite{Logginov JETP L 07}
gives an estimation for the constant
$\gamma_{0}=\sigma/\gamma^{2}=10^{-9}$ CGS.
It corresponds to $\sigma=4\times10^{4}$ CGS. 
Hence, we can estimate the polarization $P=\sigma S_{0}^{2}\tilde{k}=4$ CGS,
with $\tilde{k}\approx10^{6}$ cm$^{-1}$
(for the cycloid with period of $62$ nm \cite{ZvezdinMukhin JETP L 09}),
and
the concentration $n_{0}=10^{22}$cm$^{-3}$.
In
Ref. \cite{Sparavigna PRB 94}
constant $\gamma_{0}$ is given as $\gamma_{0}=\sigma/\gamma^{2}=10^{-11}$ CGS.
It corresponds to $\sigma=4\times10^{2}$ CGS.

Refs.
\cite{AndreevTrukh PS 24}, \cite{AndreevTrukh JETP 24}
give the microscopic derivation of equation
(\ref{MFMemf P def expanded})
together with the microscopic interpretation of the coefficient
\begin{equation}\label{MFMemf sigma estimation}
\sigma=
\frac{1}{2}
A \frac{\mid\gamma\mid}{c} \approx 0.4\times10^{2} \end{equation}
CGS,
where $A=2\times10^{5}$ cm$^{3}$/g
(corresponding to constant $A$ in Ref. \cite{Risinggard SR 16} in Table 1.),
$\mid\gamma\mid=0.6\times10^{7}$ CGS.
Here we use $\sigma=-g_{(\alpha)}/3$, $g_{(\alpha)}=\gamma g_{(u)}/(4c)$, and $A=g_{(u)}/6$.

Constant $\sigma=-g_{(\alpha)}/3$ shows relation between
the polarization $\textbf{P}$ and the spin density $\textbf{S}$
(see eq. (\ref{MFMemf P def expanded})).
While constant $\gamma_{0}=\sigma/\gamma^{2}$ shows the relation between
the polarization $\textbf{P}$ and the magnetization $\textbf{M}=\gamma \textbf{S}$
(see eq. 2 in Ref. \cite{Risinggard SR 16}).
Equation (\ref{MFMemf P def expanded}) allows to estimate the polarization based on the given value of $\sigma$.

The generalized spin-current model presented in Refs.
\cite{AndreevTrukh PS 24}, \cite{AndreevTrukh JETP 24}
is considered for the simple regime,
but gives a generalization of the spin-current model suggested earlier
\cite{Sergienko PRL 06}, \cite{Katsura PRL 05}, \cite{Sergienko PRB 06}
(see also review articles \cite{Tokura RPP 14}, \cite{Dong AinP 15}).
Basically,
authors consider the single
edge
of the elementary cell.
Hence, no feature of the crystal order has been included in the model.
Some further
detalization
of the model is required.
Nevertheless, it gives the analytical and numerical estimations of the constant
with the clarification of the mechanism of the polarization formation.
It includes the specification of the interaction responsible for this effect.
So, Ref. \cite{AndreevTrukh PS 24} shows
that the simultaneous action of the spin-orbit interaction and the Coulomb exchange interaction
(in the Heisenberg Hamiltonian form)
leads to the formation of polarization (\ref{MFMemf P def expanded})
and the effective magnetoelectric interaction
(\ref{MFMemf evol S under infl of E fin 00})-(\ref{MFMemf evol S under infl of E fin 02}).
Polarization (\ref{MFMemf P def expanded}) is nonzero for some configurations of the noncollinear spins.
While the Heisenberg exchange interaction tends to form the parallel order.
Hence, if other interactions,
like the Dzyloshinskii--Moriya interaction,
form the noncollinear spin order,
when the Heisenberg exchange interaction together with the
spin-orbit interaction forms the polarization on this background,
as the second level of this hierarchy.
While the Heisenberg exchange interaction is one of the larger interactions in the magnetic subsystems,
the simultaneous account of the small relativistic spin-orbit interaction leads to the relatively small polarization.
Let us point out the following Refs.
\cite{AndreevTrukh PS 24} and \cite{AndreevTrukh JETP 24}:
the simultaneous account of the spin-orbit interaction and
the Dzyloshinskii--Moriya interaction
leads to the formation of the electric polarization in the systems of parallel spins
$\textbf{P}\sim \textbf{S}^{2}$
(or the collinear part of the partially collinear spin gives rise to the polarization formation).

The materials with the possibility of the formation of the electric polarization symmetric relative to the spins of the magnetic ions
usually have the larger value of the polarization in comparison with multiferroic materials,
where the magnetoelectric coupling appears due to the noncollinear part of the spins
(see equations (\ref{MFMemf edm def simm}), (\ref{MFMemf edm operator Mod}) and (\ref{MFMemf P def expanded})).
It may look strange
that the stronger regime is associated with the Dzyloshinskii--Moriya interaction,
while the weaker regime is related to the Heisenberg exchange interaction
\cite{AndreevTrukh PS 24}, \cite{AndreevTrukh JETP 24}.
While the coefficient of the Heisenberg exchange interaction Hamiltonian is usually 10 times larger
than the coefficient of the Dzyloshinskii--Moriya interaction Hamiltonian.
The solution to this item follows from the symmetry of these Hamiltonians.
The Heisenberg exchange interaction being more symmetric gives the polarization in the second order of the expansion on the relative distance.
It is reflected in the form of coefficient $g_{(\alpha)}$ containing the square of the relative distance $\xi^{2}$ under the integral.
It is also related to the presence of the space derivative of the spin density in equation (\ref{MFMemf P def expanded}).
While the symmetric form of polarization $\textbf{P}\sim g_{(\beta)}\textbf{S}^{2}$
includes no space derivatives,
where $g_{(\beta)}=\int \beta(\xi)d^{3}\xi$ and
$\beta(\xi)$ is the part of the Dzyloshinskii constant presented in the
Keffer
form $\textbf{D}=\beta(\xi)\mbox{\boldmath $\delta$}\times\mbox{\boldmath $\xi$}$,
with the vector describing the shift of the nonmagnetic ion $\mbox{\boldmath $\delta$}$.
Hence, the form of the coefficient and the presence of the space derivatives demonstrate
the weakening of the contribution of the stronger Heisenberg exchange interaction.
It opens up the possibility of the Dzyloshinskii--Moriya interaction to create the larger electric polarization.

\section{Dielectric permeability}

The final Landau--Lifshitz--Gilbert equation applied in this paper has the following form:
$$\partial_{t}\textbf{S}=
\gamma[\textbf{S}\times\textbf{B}]
+A[\textbf{S}\times\triangle\textbf{S}]
+\kappa [\textbf{S}\times S_{z}\textbf{e}_{z}] $$
$$-\sigma
\biggl[ [\textbf{E}\times \nabla] S^{2}
-2(\textbf{S}\cdot[\textbf{E}\times\nabla]) \textbf{S}$$
\begin{equation}\label{MFMemf s evolution}
-S^{2}(\nabla\times\textbf{E})
+\textbf{S}(\textbf{S}\cdot [\nabla\times\textbf{E}])
\biggr]
+a[\textbf{S}\times\partial_{t}\textbf{S}]. \end{equation}
The right-hand side of equation (\ref{MFMemf s evolution})
contains the following interactions:
the magnetic dipole-dipole interaction
(including the interaction of the magnetic moments with the external magnetic field),
the mean-field limit of the Coulomb exchange interaction in the form of Heisenberg Hamiltonian,
the torque related to the anisotropy energy,
the electric dipole-dipole interaction
(including the interaction of the electric moments with the external electric field),
given by the group of terms proportional to $\sigma$,
the Gilbert damping, correspondingly.
Coefficients have the following signs:
$\kappa>0$, $A>0$, $\sigma>0$, $a<0$, $\gamma<0$.
Following the microscopic theory,
we can present the coefficient
$A=\frac{1}{6}g_{u}$,
where $g_{u}$ is the second moment of the function describing the space dependence of the potential energy in the Heisenberg Hamiltonian.

We consider perturbations of the spin density together with the perturbations of the electromagnetic field,
so we use
the Maxwell's equations
\begin{equation}\label{MFMemf rot E} \nabla\times \textbf{E}=-\frac{1}{c}\partial_{t}\textbf{B},\end{equation}
and
\begin{equation}\label{MFMemf rot B}
\nabla\times \textbf{B}=\frac{1}{c}\partial_{t}\textbf{E} +4\pi\nabla\times \textbf{M}+\frac{4\pi}{c}\partial_{t}\textbf{P}, \end{equation}
where we represent the magnetization via the spin density
$\textbf{M}=\gamma\textbf{S}$ and include the gyromagnetic ratio $\gamma$,
and zero current density $\textbf{j}=0$.

The first three terms in the Landau--Lifshitz--Gilbert equation lead to
the characteristic frequency of the magnetically ordered materials
$\Omega\equiv \gamma B_{0} +A k^2 S_0 +\kappa S_{0}$
(here we have $S_{0z}=S_0>0$, $B_{0z}=B_0<0$, and $\gamma B_{0}>0$).
The last term, describing relaxation,
modifies this characteristic frequency,
adding the imaginary part
$\Omega_{0}=\Omega+\imath a\omega S_{0}$.
The group of terms proportional to constant $\sigma$ gives two contributions.
One of them is related to the perturbations of the electric field.
Its contribution is discussed below,
after the derivation of the dielectric permeability.
However, another group related to the constant electric field.
It gives the characteristic frequency
$\Sigma=\Sigma_{x}+\Sigma_{y}=2S_{0}\sigma (E_{0y}k_{x}-E_{0x}k_{y})$.

We consider the following structure of the spin density, electric and magnetic field:
$\textbf{S}=\textbf{S}_{0}+\delta \textbf{S}$,
$\textbf{B}=\textbf{B}_{0}+\delta \textbf{B}$,
$\textbf{E}=\textbf{E}_{0}+\delta \textbf{E}$,
where the perturbations of the spin density have the following form
$\delta \textbf{S}=\textbf{s}_{a}e^{-\imath\omega t+\imath \textbf{k}\textbf{r}}$
with the complex amplitude $\textbf{s}_{a}$.
Other perturbations have similar structures.

For the chosen equilibrium state and described form of the perturbations,
we obtain the linearized form
of the Landau--Lifshitz--Gilbert equation
for the small amplitude perturbations
in order to detect the tendency to the instability of the system of parallel spins due to the
magnetoelectric coupling
at the propagation of the electromagnetic waves in the dielectric material:
\begin{equation}\label{MFMemf S x eq for pert}
-\imath(\omega-\Sigma) \delta S_{x} -\Omega_{0}\delta S_{y}
= \imath\sigma S_{0}^{2} (k_{y}\delta E_{z}-k_{z}\delta E_{y})-\gamma S_{0}\delta B_{y}, \end{equation}
and
\begin{equation}\label{MFMemf S y eq for pert}
-\imath(\omega-\Sigma) \delta S_{y} +\Omega_{0}\delta S_{x}
= \imath\sigma S_{0}^{2} (k_{z}\delta E_{x}-k_{x}\delta E_{z})+\gamma S_{0}\delta B_{x}. \end{equation}

Equilibrium polarization is equal to zero in the considered regime $\textbf{P}_{0}=0$.
However, we consider the perturbations of the polarization caused by the magneto-electric effect:
\begin{equation}\label{MFMemf P pert via S}
\delta\textbf{P}=\sigma [\textbf{S}_{0} (\nabla\cdot \delta\textbf{S})
- (\textbf{S}_{0}\cdot \nabla)\delta\textbf{S}]. \end{equation}

The linearized form of the
Landau--Lifshitz--Gilbert equation
(\ref{MFMemf S x eq for pert}) and (\ref{MFMemf S y eq for pert})
shows
that the spin density perturbations can be considered as the function of the electric and magnetic fields
\begin{equation}\label{MFMemf }
\delta S^{\alpha}=X^{\alpha\beta}\delta B^{\beta} +Y^{\alpha\beta}\delta E^{\beta}, \end{equation}
however, we can represent the electric field perturbations
via the magnetic field perturbations or
vice versa.
Presented form of the spin density leads to the analogous form of the polarization (\ref{MFMemf P pert via S})
\begin{equation}\label{MFMemf P pert via E and B}
\delta P^{\alpha}=\kappa^{\alpha\beta}\delta E^{\beta} +Z^{\alpha\beta}\delta B^{\beta}. \end{equation}

\subsection{On the definition of the dielectric permeability}

To avoid some
misunderstanding,
we discuss here three definitions for the dielectric permeability.

Let us start with the most simple definition
\begin{equation}\label{MFMemf susceptibility 1}
\varepsilon^{\alpha\beta}_{1}=\delta^{\alpha\beta}+4\pi\kappa^{\alpha\beta}, \end{equation}
where
$\delta^{\alpha\beta}$ is the three-dimensional Kronecker symbol
presenting the unit matrix.

The second step is the excluding of the magnetic field from equation (\ref{MFMemf P pert via E and B})
in order to obtain the generalized form of the dielectric
susceptibility
\begin{equation}\label{MFMemf susceptibility 2}
\varepsilon^{\alpha\beta}_{2}=\delta^{\alpha\beta}
+4\pi\biggl(\kappa^{\alpha\beta} +Z^{\alpha\mu}\frac{k^{\nu}c}{\omega}\varepsilon^{\mu\nu\beta}\biggr). \end{equation}

The third definition based on the structure of the
right-hand side of equation (\ref{MFMemf rot B})
\begin{equation}\label{MFMemf eq for susceptibility 3}
\delta E^{\alpha}+4\pi\delta P^{\alpha}-\frac{4\pi\gamma c}{\omega}\varepsilon^{\alpha\mu\nu}k^{\mu}\delta S^{\nu}
\equiv\varepsilon^{\alpha\beta}_{3}\delta E^{\beta}. \end{equation}
Equation (\ref{MFMemf eq for susceptibility 3}) leads to most complex of three given expressions
$$\varepsilon^{\alpha\beta}_{3}=\delta^{\alpha\beta}
+4\pi\kappa^{\alpha\beta}
+Z^{\alpha\gamma}\varepsilon^{\beta\gamma\mu}\frac{k^{\mu}c}{\omega}
-4\pi\gamma \varepsilon^{\alpha\mu\nu}Y^{\nu\beta}\frac{k^{\mu}c}{\omega} $$
\begin{equation}\label{MFMemf susceptibility 3}
-4\pi\gamma \varepsilon^{\alpha\mu\nu}\frac{k^{\mu}c}{\omega} \frac{k^{\lambda}c}{\omega}
X^{\nu\sigma}\varepsilon^{\sigma\lambda\beta} . \end{equation}
Equation (\ref{MFMemf susceptibility 3}) is applied below.

Let us point out the definition of
the refractive index used in this paper
$n\equiv kc/\omega$.

\section{Waves propagating perpendicular to the anisotropy axis}


We consider the regime of $\textbf{k}=\{ k_{x},0,0\}$,
where there are two characteristic frequencies
$\Omega_{s}\equiv \mid\gamma\mid c/(\sigma S_{0})$
and
$\Sigma=\Sigma_{x}=2S_{0}\sigma E_{0y}k_{x}$.

This regime leads to the following expression for the refractive index
via the single element of the dielectric permeability tensor
\begin{equation}\label{MFMemf n2 with epsilon zz}n^{2}=\varepsilon_{zz}.\end{equation}

We obtain the explicit form of the dielectric permeability
\begin{equation}\label{MFMemf varepsilon_zz}
\varepsilon_{zz}=1-\frac{8\pi \sigma S_{0}^{2}\mid\gamma\mid}{c}\frac{k_{x}^{2}c^{2}}{\Lambda}
\biggl[1-\frac{\Sigma}{\omega}-\frac{\Omega_{0}}{\Omega_{s}}\biggr], \end{equation}
where
$\Lambda\equiv\Omega_{0}^{2}-(\omega-\Sigma)^{2}$.
We also get an extra characteristic dimensionless parameter
$\epsilon=8\pi \sigma S_{0}^{2}\mid\gamma\mid/c$.
We point out that frequency $\Sigma$ can be positive $\Sigma>0$
(if vectors $\textbf{k}$, $\textbf{E}_{0}$, and $\textbf{S}_{0}$ form the right vector triple
i.e. $\textbf{k}\times\textbf{E}_{0}=+\textbf{S}_{0}$)
or negative $\Sigma<0$.

Parameter $\frac{\Omega_{0}}{\Omega_{s}}$ can be comparable with 1
if we consider relatively large wave vectors $k$.
Overwise, it can be dropped in $\varepsilon_{zz}$.
The dimensionless parameter $\frac{\Sigma}{\omega}$ depends on the frequency.
However, if we consider frequency $\omega$ of order of $10^{10}$ s$^{-1}$ we find $\frac{\Sigma}{\omega}\ll 1$.
Similarly, frequency $\Sigma$ can be dropped in $\Lambda$.
All of it simplifies the expression for the dielectric permeability
\begin{equation}\label{MFMemf varepsilon_zz simpl via kc}
\varepsilon_{zz,simplified}=1-\frac{8\pi \sigma S_{0}^{2}\mid\gamma\mid}{c}
\frac{k_{x}^{2}c^{2}}{\Omega_{0}^{2}-\omega^{2}}. \end{equation}
This expression shows the explicit dependence of the dielectric permeability
on the refractive index $n^{2}\sim k_{x}^{2}c^{2}$.
Let us explicitly demonstrate this dependence
\begin{equation}\label{MFMemf varepsilon_zz simpl via n}
n^{2}=1-\epsilon n^{2}
\frac{\omega^{2}}{\Omega_{0}^{2}-\omega^{2}}, \end{equation}
where $\Omega_{0}$ is also a function of the wave vector,
but this part is not interpreted via the refractive index.
Finally, we get the following expression for the refractive index as the function of real frequency
$n^{2}(\omega)$
\begin{equation}\label{MFMemf n via freq}
n^{2}=\frac{1}{1+\epsilon\frac{\omega^{2}}{\Omega_{0}^{2}-\omega^{2}}}. \end{equation}

Both equations for the dielectric permeability
(\ref{MFMemf varepsilon_zz}) and (\ref{MFMemf varepsilon_zz simpl via kc})
show that its imaginary part $\varepsilon_{zz}=Re(\varepsilon_{zz})+\imath Im(\varepsilon_{zz})$
is negative $Im(\varepsilon_{zz})<0$.
It can demonstrate the presence of an instability.

Equation (\ref{MFMemf n via freq}) shows the negative value of the square of the refractive index
$n^{2}=Re n^{2}+\imath Im n^{2}$, with $Im n^{2}<0$.
It also leads to the negative value of the refractive index itself
since
$n=+\sqrt{n^{2}}=+Re n \cdot (1+\imath Im n^{2}/(2Re n^{2}))$,
$n=Re n+\imath Im n$
and
$sign (Im n)=sign (Im n^{2})$.
We get it along with
$\delta \textbf{S}\sim e^{-\imath\omega t+\imath k_{x}x}$
$\sim e^{+\imath k_{0x}nx}\sim e^{- k_{0x} (Imn)x}$,
where $k_{0x}$ is the projection of the wave vector in vacuum for the electromagnetic wave.
So, we get the increase of the amplitude in the small amplitude regime.

Dispersion equation
(\ref{MFMemf n2 with epsilon zz})
does not show the dispersion dependence of the spin waves propagating perpendicular to the anisotropy axis explicitly.
However, we can extract the required solution.
To this end, we need to consider the limit $\sigma\rightarrow0$.
It allow us to consider the limit $\Lambda\rightarrow0$,
where
$\omega^{2}\approx\Omega_{0}^{2}$,
while we drop the contribution of $\Sigma$.
The contribution of the Zeeman energy/dipole-dipole interaction is anisotropic.
It disappears in this regime,
while the method derivation of the dispersion dependence via the Maxwell equations is bound to the dipole-dipole interaction.
Therefore, the extraction of the spin wave dispersion dependence requires additional efforts.
While, in this simple model, the spin waves do not affect the propagation of the electromagnetic waves in the chosen direction.

\subsection{Dispersion dependence and instability}

Next, we consider the dispersion dependence obtained from equation (\ref{MFMemf varepsilon_zz simpl via n})
in order to
identify the instabilities
$$\omega_{\pm}^{2}(k_{x})=\frac{1}{2}\biggl[\Omega_{0}^{2}+k_{x}^{2}c^{2}(1-\epsilon)$$
\begin{equation}\label{MFMemf omega pm 1}
\pm\sqrt{(\Omega_{0}^{2}-k_{x}^{2}c^{2})^{2}
-2\epsilon k_{x}^{2}c^{2}\Omega_{0}^{2} +\epsilon^{2}k_{x}^{4}c^{4}}\biggr], \end{equation}
where the complex frequency $\omega$ is the function of the real wave vector $k_{x}$.

Start our analysis with the regime of large value of $\Omega^{2}-k_{x}^{2}c^{2}$.
Present corresponding representation of equation (\ref{MFMemf omega pm 1}):
$$\omega_{\pm}^{2}(k_{x})=\frac{1}{2}\biggl[\Omega_{0}^{2}+k_{x}^{2}c^{2}$$
\begin{equation}\label{MFMemf omega pm 2}
\pm (k_{x}^{2}c^{2}-\Omega_{0}^{2})
\sqrt{1-\frac{2\epsilon k_{x}^{2}c^{2}\Omega_{0}^{2}}{(k_{x}^{2}c^{2}-\Omega_{0}^{2})^2}}\biggr]. \end{equation}

It leads to the approximate forms of the dispersion dependencies:
\begin{equation}\label{MFMemf omega p 2}
\omega_{+}^{2}=k_{x}^{2}c^{2}\biggl(
1+\frac{\epsilon\Omega_{0}^{2}}{\Omega_{0}^{2}-k_{x}^{2}c^{2}}
\biggr), \end{equation}
and
\begin{equation}\label{MFMemf omega m 2}
\omega_{-}^{2}=\Omega_{0}^{2}\biggl(
1-\frac{\epsilon k_{x}^{2}c^{2}}{\Omega_{0}^{2}-k_{x}^{2}c^{2}}
\biggr), \end{equation}
where
$\Omega_{0}^{2}=
\Omega^{2}
- a^{2} S_{0}^{2}\omega^{2}
+2\imath a S_{0}\omega\Omega$,
with $\omega\approx\omega_{\pm}|_{\epsilon=0}$.

To identify the instability we need to consider the imaginary part of the obtained dispersion dependencies.
Positive part of the imaginary frequency is found for $\omega_{+}$ branch
\begin{equation}\label{MFMemf omega Im p 2}
Im\omega_{+}=-2a S_{0}
\frac{\epsilon\Omega k_{x}^{4}c^{4}}{(\Omega^{2}-k_{x}^{2}c^{2})^2 +4a^2 S_0^2 \Omega^{2}k_{x}^{2}c^{2}}>0
, \end{equation}
showing the increase of the amplitude.
This tendency found for the small amplitude regime.
Further evolution of the amplitude leads to the formation of noncollinear equilibrium spin structure.

Next we consider the regime of small value of $\Omega^{2}-k_{x}^{2}c^{2}$,
so $\Omega_{0}^{2}-k_{x}^{2}c^{2}\approx 2\imath aS_{0}\omega\Omega$.
We also include $\omega\approx\Omega$.
Equation (\ref{MFMemf omega pm 1}) can be represented in the following form
\begin{equation}\label{MFMemf }
\omega_{\pm}^{2}=\Omega^{2}\biggl[1-\imath\mid \tilde{a}\mid\pm\imath\sqrt{\tilde{a}^{2}+\frac{1}{2}\epsilon}\biggr], \end{equation}
with the dimensionless parameter $\tilde{a}=aS_{0}$.
Frequency $\omega_{-}^{2}$ shows small increase of the damping related to the dielectric effects $\epsilon>0$.
However, $\omega_{+}^{2}$ demonstrates the instability related to the positive value of constant $\epsilon>0$:
\begin{equation}\label{MFMemf }
\frac{\omega_{+}^{2}}{\Omega^{2}}=1+\frac{1}{2}\imath\frac{\epsilon}{\mid \tilde{a}\mid+\sqrt{\tilde{a}^{2}+\frac{1}{2}\epsilon}}.
\end{equation}
Assuming $\omega_{+}\approx\Omega(1+\imath\xi)$
we find
\begin{equation}\label{MFMemf }
\xi=\frac{1}{4}\frac{\epsilon}{\mid \tilde{a}\mid+\sqrt{\tilde{a}^{2}+\frac{1}{2}\epsilon}}.
\end{equation}
Corresponding perturbation of the spin density can be presented as
$\delta \textbf{S}\sim e^{-\imath\omega t}$$\sim e^{\Omega\xi t}$.
Here we present an
estimation of dimensionless parameter
$\epsilon=8\pi \sigma S_{0}^{2}\mid\gamma\mid/c \approx 2\times10^{-8}$
for $\sigma=4\times10^{4}$ CGS, $S_{0}\approx\hbar n_{0}=10^{-5}$ g/(cm s),
$\gamma=-0.6\times10^{7}$ CGS.
It gives $\xi\approx 10^{-6}$ at $aS_{0}=0.01$.

Noncollinear order of the spins can affect the relative positions of the magnetic and nonmagnetic ions.
The spin wave perturbations create oscillations of the directions of the magnetic moments.
As the result, the oscillating electric dipole moment appears in the medium.
The action of the electric field
on the electric dipoles associated with the magnetic moments
creates the torque acting on the magnetic moments.
Perturbations of the electric field associated with the propagation of the electromagnetic wave contribute in the described torque.
One of the circularly polarized electromagnetic eigen-waves is coupled to the spin wave
(at the propagation parallel to the anisotropy axis).
However, the linearly polarized electromagnetic wave is not coupled to the spin wave
(at the propagation perpendicular to the anisotropy axis)
in the simple models of the magnetically ordered dielectrics.
The instability found in Sec. IV appears as the result of the coupling
of the linearly polarized electromagnetic wave to the spin wave.

\section{Waves propagating parallel to the anisotropy axis}

The standard regime for the spin waves and the electromagnetic waves in the magnetic materials appears in the presented model
if we consider the waves propagating parallel to the anisotropy axis
$\textbf{k}=\{ 0,0,k_{z}\}$.
In this regime we find one additional characteristic frequency
$\Omega_{s}\equiv \mid\gamma\mid c/(\sigma S_{0})$
(while $\Sigma=0$)
associated with the magnetoelectric coupling.

We obtain two branches of the dispersion dependencies for the circularly polarized waves with different state of polarization:
$$n_{-}^{2}=
\varepsilon_{xx}-\imath\varepsilon_{xy} $$
\begin{equation}\label{MFMemf n2-}
= 1+ \frac{4\pi\gamma^{2}S_{0}}{\Omega_{0}-\omega}
\biggl(\frac{k_{z}c}{\omega}\biggr)^{2}\biggl(1-\frac{\omega}{\Omega_{s}}\biggr)^{2}
, \end{equation}
and
$$n_{+}^{2}=
\varepsilon_{xx}+\imath\varepsilon_{xy}$$
\begin{equation}\label{MFMemf n2+}
= 1+ \frac{4\pi\gamma^{2}S_{0}}{\Omega_{0}+\omega}
\biggl(\frac{k_{z}c}{\omega}\biggr)^{2}\biggl(1+\frac{\omega}{\Omega_{s}}\biggr)^{2}. \end{equation}
Equation (\ref{MFMemf n2-}) contains the contribution of both the electromagnetic wave and the spin wave,
while equation (\ref{MFMemf n2+}) contains the contribution of the electromagnetic wave only.

An interesting manifestation of the magneto-electric effect can be found in equation
(\ref{MFMemf n2-}),
where we find
that the refractive index goes to 1, so $n\rightarrow1$ at $\omega\rightarrow \Omega_{s}$.
Using $\sigma=4\times10^{4}$ CGS (following Ref. \cite{Logginov JETP L 07})
we get an estimation of $\Omega_{s}\sim 10^{17}$ s$^{-1}$.
It corresponds to the transparency of the medium (its magnetic part).
Existing small values of polarization lead to high frequency in the strong ultraviolet diapason.
Increase of the polarization may shift $\Omega_{s}$ down to
the visible range $\sim10^{15}$ s$^{-1}$.

In both cases we find $Im (\varepsilon_{xx}\pm\imath\varepsilon_{xy}) >0$
(since $a<0$),
hence we expect stability of the chosen equilibrium relatively perturbations propagating in the chosen direction.

The right-hand side of equations
(\ref{MFMemf n2-}) and (\ref{MFMemf n2+})
contains the square of
refractive index $n=k_{z}c/\omega$:
\begin{equation}\label{MFMemf }
n_{\pm}^{2}= \frac{1}{1- \frac{4\pi\gamma^{2}S_{0}}{\Omega_{0}\pm\omega}
(1\pm\frac{\omega}{\Omega_{s}})^{2}}
, \end{equation}
with $Im n_{\pm}^{2} >0$,
which corresponds to the stability,
like it follows from the dielectric permeability (\ref{MFMemf n2-}) and (\ref{MFMemf n2+}).

\section{Conclusion}

No Dzylaoshinskii-Moriya interaction has been considered in the applied model for the explicit demonstration
that the interaction of the electric dipole moments with the electric field causes the instability of the collinear order of the spins,
while it is frequently associated with Dzylaoshinskii-Moriya interaction to form any noncollinear order of spins.
The obtained instability demonstrates itself via the imaginary part of the frequency existing in spite of the
Gilbert damping.
Hence, there is the growth of the spin deviation from the equilibrium state
towards the formation of another equilibrium state in the nonlinear regime.
It is also demonstrated
that the instability reveals itself in the sign of the imaginary part of the refractive index and the elements
dielectric permeability tensor.
Let us point out
that the
instability
has been found for the collinear spins lined up in parallel to the anisotropy axis
in the easy-axis ferromagnetic material,
where the linearly polarized wave propagates in the direction perpendicular to the anisotropy axis
(the electric field in this wave is parallel to the anisotropy axis).

\section{DATA AVAILABILITY}

Data sharing is not applicable to this article as no new data were
created or analyzed in this study, which is a purely theoretical one.

\section{Acknowledgements}

The work is supported by the Russian Science Foundation under the
grant
No. 25-22-00064.


\begin{thebibliography}{17}



\bibitem{Camley SSR 23}
R. E. Camley, K. L. Livesey
"Consequences of the Dzyaloshinskii-Moriya interaction",
Surface Science Reports \textbf{78}, 100605 (2023).

\bibitem{Pyatakov UFN 12} A. P. Pyatakov and A. K. Zvezdin,
"Magnetoelectric and multiferroic media",
Phys.-Usp. \textbf{55}, 557 (2012).

\bibitem{Khomskii JETP 21}
D. I. Khomskii
"Multiferroics and Beyond: Electric Properties of Different Magnetic Textures",
Journal of Experimental and Theoretical Physics \textbf{132}, 482 (2021).


\bibitem{Risinggard SR 16} V. Risinggard, I. Kulagina, J. Linder,
"Electric field control of magnoninduced magnetization dynamics in multiferroics",
Scientific Reports \textbf{6}, 31800 (2016).

\bibitem{ZvezdinMukhin JETP L 09} A. K. Zvezdin, A. A. Mukhin,
"On the effect of inhomogeneous magnetoelectric (flexomagnetoelectric) interaction on the
spectrum and properties of magnons in multiferroics",
JETP Lett. \textbf{89}, 328–332, (2009).

\bibitem{Mostovoy npj 24} M. Mostovoy,
"Multiferroics: different routes to magnetoelectric coupling",
npj Spintronics, 2:18 (2024)

\bibitem{Tokura RPP 14} Y. Tokura, S. Seki, and N. Nagaosa,
"Multiferroics of spin origin",
Rep. Prog. Phys. \textbf{77}, 076501 (2014).


\bibitem{AndreevTrukh PS 24} P. A. Andreev, M. I. Trukhanova,
"Electric polarization evolution equation for antiferromagnetic
multiferroics with the polarization proportional to the scalar product
of the spins",
Phys. Scr. \textbf{99}, 1059b2 (2024).

\bibitem{AndreevTrukh EPJ B 24} P. A. Andreev, M. I. Trukhanova,
"Polarization evolution equation for exchange-strictionally formed type II multiferroic materials",
Eur. Phys. J. B \textbf{97}, 116 (2024).



\bibitem{AndreevTrukh JETP 24} P. A. Andreev, M. I. Trukhanova,
"Equation of evolution of electric polarization of multiferroics proportional to the vector
product of spins of ions of the cell under the influence of the Heisienberg Hamiltonian",
JETP, 2024, Vol. \textbf{166}, Issue 5 (11), pp. 665–678, 2024 [in russian].


\bibitem{Andreev Ph B 24}
P. A. Andreev,
"Hydrodynamic model of skyrmions and vorticities in the spin-1 Bose–Einstein condensate in ferromagnetic phase at finite temperatures",
Physica B: Condensed Matter \textbf{695}, 416470 (2024).


\bibitem{Andreev 23 11}
M. I. Trukhanova, P. A. Andreev, Y. N. Obukhov,
"A new quantum hydrodynamic description of ferroelectricity in spiral magnets",
International Journal of Modern Physics B, 2550072 (2024).


\bibitem{Sparavigna PRB 94}
A. Sparavigna, A. Strigazzi, and A. Zvezdin,
"Electric-field effects on the spin-density wave in magnetic ferroelectrics",
Phys. Rev. B \textbf{50}, 2953 (1994).

\bibitem{Mostovoy PRL 06} M. Mostovoy,
"Ferroelectricity in Spiral Magnets",
Phys. Rev. Lett. \textbf{96}, 067601 (2006).

\bibitem{Dong AinP 15}
S. Dong, J.-M. Liu, S.-W. Cheong, Z. Ren,
"Multiferroic materials and magnetoelectric physics: symmetry, entanglement, excitation, and topology",
Advances in Physics \textbf{64}, 519 (2015).





\bibitem{Logginov JETP L 07} A. S. Logginov, G. A. Meshkov, A. V. Nikolaev, A. P. Pyatakov,
"Magnetoelectric control of domain walls in a ferrite garnet film",
JETP Lett. \textbf{86}, 115–118, (2007).




\bibitem{Sergienko PRL 06} I. A. Sergienko, C. Sen, and E. Dagotto,
"Ferroelectricity in the Magnetic E-Phase of Orthorhombic Perovskites", Phys. Rev. Lett. \textbf{97}, 227204 (2006).


\bibitem{Katsura PRL 05} H. Katsura, N. Nagaosa, and A. V. Balatsky,
"Spin Current and Magnetoelectric Effect in Noncollinear Magnets", Phys. Rev. Lett. \textbf{95}, 057205 (2005).

\bibitem{Sergienko PRB 06} I. A. Sergienko, E. Dagotto,
"Role of the Dzyaloshinskii-Moriya interaction in multiferroic perovskites", Phys. Rev. B \textbf{73}, 094434 (2006).





\end{thebibliography}
\end{document}